\documentclass[aps,prl,floatfix,superscriptaddress,reprint]{revtex4-1} 
\usepackage{graphicx}
\usepackage{amsmath,amsfonts,amssymb}
\usepackage{xcolor}
\usepackage{pdfpages}

\makeatletter
\AtBeginDocument{\let\LS@rot\@undefined}
\makeatother

\DeclareMathOperator{\sign}{sign}

\begin{document}

\title{Spin Vortex Crystal Order in Organic Triangular Lattice Compound}
\date{\today}
\begin{abstract}  Organic salts represent an ideal experimental playground for studying the interplay between magnetic and charge degrees of freedom, which has culminated in the discovery of several spin-liquid candidates, such as $\kappa$-(ET)$_2$Cu$_2$(CN)$_3$ ($\kappa$-Cu). Recent theoretical studies indicate the possibility of chiral spin liquids stabilized by ring-exchange, but the parent states with chiral magnetic order have not been observed in this material family. In this work, we discuss the properties of the recently synthesized $\kappa$-(BETS)$_2$Mn[N(CN)$_2$]$_3$ ($\kappa$-Mn). Based on analysis of specific heat, magnetic torque, and NMR measurements combined with {\it ab initio} calculations, we identify a spin-vortex crystal order. These observations definitively confirm the importance of ring-exchange in these materials, and support the proposed chiral spin-liquid scenario for triangular lattice organics. \end{abstract}

\author{Kira Riedl}
\affiliation{Institut f\"ur Theoretische Physik, Goethe-Universit\"at Frankfurt,
Max-von-Laue-Strasse 1, 60438 Frankfurt am Main, Germany}
\author{Elena Gati}
\affiliation{Physikalisches Institut, Goethe-Universit\"at Frankfurt,  Max von Laue Str 1, 60438 Frankfurt am Main, Germany}
\affiliation{Max Planck Institute for Chemical Physics of Solids, Dresden, Germany}
\author{David Zielke}
\affiliation{Physikalisches Institut, Goethe-Universit\"at Frankfurt,  Max von Laue Str 1, 60438 Frankfurt am Main, Germany}
\author{Steffi Hartmann}
\affiliation{Physikalisches Institut, Goethe-Universit\"at Frankfurt,  Max von Laue Str 1, 60438 Frankfurt am Main, Germany}
\author{Oleg M. Vyaselev}
\affiliation{Institute of Solid State Physics, Russian Academy of Sciences, 142432 Chernogolovka, Russia}
\author{Nataliya D. Kushch}
\affiliation{Institute of Problems of Chemical Physics, Russian Academy of Sciences, 142432 Chernogolovka, Russia}
\author{Harald O. Jeschke}
\affiliation{Research Institute for Interdisciplinary Science, Okayama University, Okayama 700-8530, Japan}
\author{Michael Lang}
\affiliation{Physikalisches Institut, Goethe-Universit\"at Frankfurt,  Max von Laue Str 1, 60438 Frankfurt am Main, Germany}
\author{Roser Valent{\'\i}}
\affiliation{Institut f\"ur Theoretische Physik, Goethe-Universit\"at Frankfurt,
Max-von-Laue-Strasse 1, 60438 Frankfurt am Main, Germany}
\author{Mark V. Kartsovnik}
\affiliation{Walther-Meissner-Institut, Bayerische Akademie der Wissenschaften, Walther-Meissner-Strasse 8, Garching D-85748, Germany}
\author{Stephen M. Winter*}
\affiliation{Department of Physics and Center for Functional Materials, Wake Forest University, NC 27109, USA}

\maketitle

{\it Introduction $-$} The role of higher order magnetic couplings in organic quantum spin-liquid (QSL) candidates such as $\kappa$-(ET)$_2$Cu$_2$(CN)$_3$ ($\kappa$-Cu) has been well discussed over the last two decades \cite{motrunich2005variational,lee2005gauge,block2011spin,holt2014spin,zhou2017quantum}. These materials are Mott insulators, but exist on the verge of itinerancy, such that conventional  nearest-neighbor magnetic couplings are insufficient to describe their low-energy spin Hamiltonian. Instead, one must consider higher order terms, which appear, for example,  in the form of four-spin ring-exchange $K_{ijkl}(\mathbf{S}_i\cdot\mathbf{S}_j)(\mathbf{S}_k\cdot\mathbf{S}_l)$. These interactions are thought to play a crucial role in destabilizing conventional magnetic orders, thus promoting a QSL ground state \cite{motrunich2005variational,holt2014spin,block2011spin}. However, {\it classical} ring-exchange models \cite{roger1989cyclic,chubukov1992phase,kubo1997ground,kubo1998possible,kubo2003magnetic,lauchli2005two,hayami2017effective,paul2020role} often feature additional unconventional ordered phases -- as first highlighted in the context of solid $^3$He \cite{roger1980four, roger1983magnetism, roger1990coupled}. Of particular interest are commensurate, non-collinear phases characterized by scalar chiral (e.g.~$\mathbf{S}_i\cdot(\mathbf{S}_j \times \mathbf{S}_k))$ and/or vector chiral (e.g. $\sum \mathbf{S}_i\times \mathbf{S}_j$) order parameters, which tend to be selected by large ring-exchange.
Indeed, a flurry of recent proposals \cite{szasz2020chiral,chen2021quantum,szasz2021phase,wietek2021mott,cookmeyer2021four} have identified a chiral spin-liquid (CSL) derived from these classical orders as a leading candidate for the ground state of $\kappa$-Cu. 
This raises an important qualifying  question; if $\kappa$-Cu is indeed a CSL, then the related chiral magnetic orders (predicted to persist away from the CSL) should also be observable in other organic materials with suitably tuned couplings.

\begin{figure}[t]
\includegraphics[width=\linewidth]{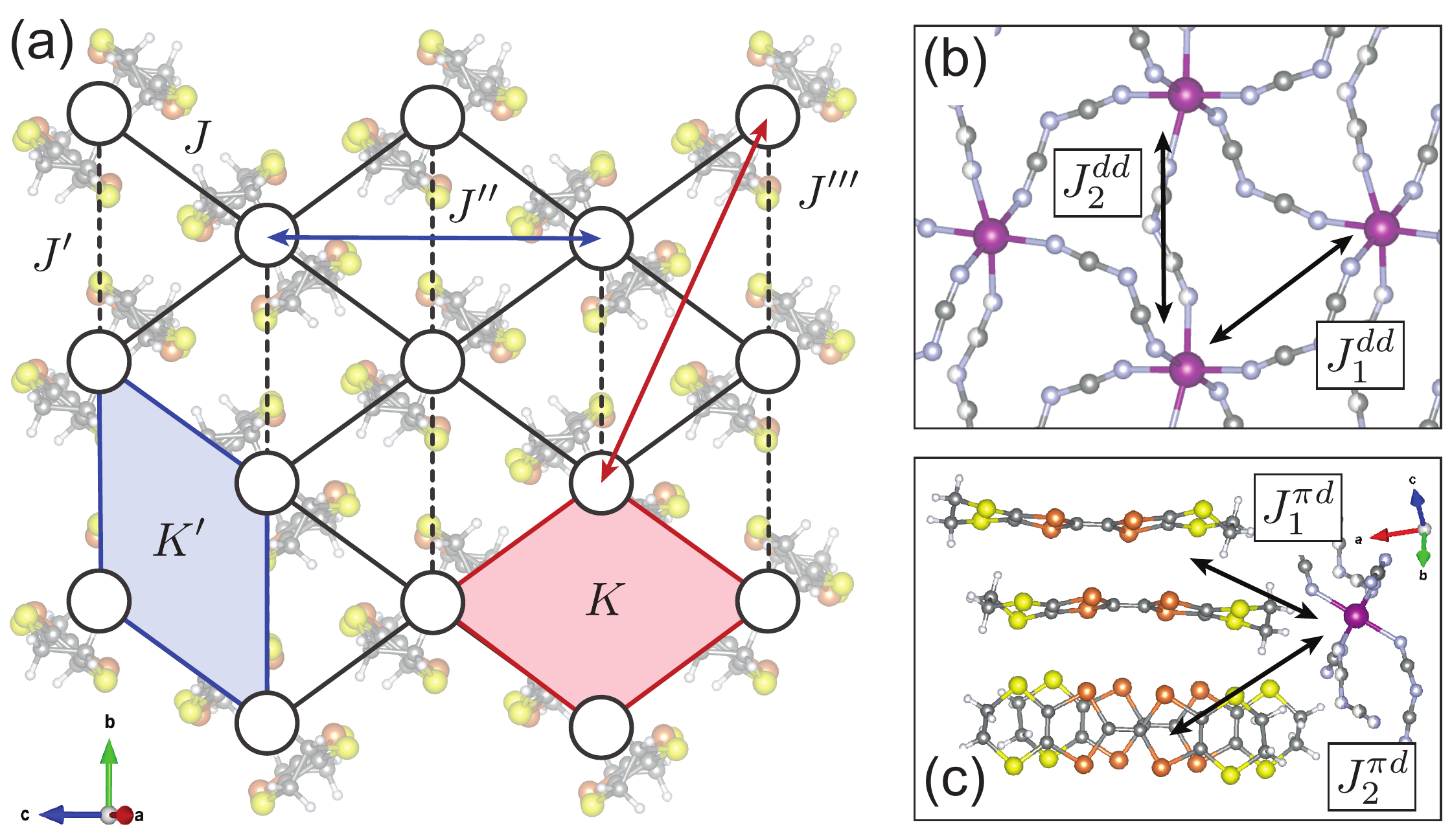}
\caption{Definition of (a) $\pi$-$\pi$ couplings between $S$=$\frac{1}{2}$ BETS dimers in the organic layer of $\kappa$-Mn,  (b) $d$-$d$ couplings between $S=\frac{5}{2}$ Mn(II) in the anion layer, (c) $\pi$-$d$ couplings between BETS and Mn. $J$ labels two-spin interactions, while $K$ labels four-spin interactions.}
\label{fig:structure}
\end{figure}

In this letter, we consider the magnetic ground state of \mbox{$\kappa$-(BETS)$_2$Mn[N(CN)$_2$]$_3$} ($\kappa$-Mn) \cite{kushch2007molecular,kushch2008pi,morgunov2007spin,PhysRevB.82.155123,PhysRevB.83.094425,vyaselev2012properties,vyaselev2012staggered,vyaselev2017interplay,zverev2019fermi}, which we demonstrate to lie in a parameter region conducive to chiral magnetic order. 
This material has a layered structure (Fig.~\ref{fig:structure}), typical of $\kappa$-phase materials \cite{Toyota07}. The organic layer is composed of [BETS]$_2^{1+}$ (= bisethylenedithio-tetraselenafulvalene) dimers forming a distorted triangular lattice with an $S=1/2$ spin per dimer. The anion layer is composed of Mn(II) ($S=5/2$) ions also forming a distorted triangular lattice, linked by dicyanamide bridges. The phase diagram of $\kappa$-Mn \cite{PhysRevB.82.155123} is similar to $\kappa$-phase ET salts such as the spin-liquid candidate $\kappa$-Cu \cite{Kurosaki05,Pustogow18} and the antiferromagnet \mbox{$\kappa$-(ET)$_2$Cu[N(CN)$_2$]$_2$Cl} \cite{Lefebvre00,Limelette03,Gati16}  ($\kappa$-Cl). All have insulating ground states at low pressure, which are suppressed in favor of metallicity/superconductivity under mild pressure. 

At elevated temperatures, $\kappa$-Mn displays a metallic temperature dependence of the electrical conductivity. Magnetic order in the BETS layer onsets at $T_{\rm N} \sim 22$ K in conjunction with a metal-insulator transition (MIT). This transition is marked by a significant broadening of the $^{13}$C NMR resonances \cite{vyaselev2012staggered,vyaselev2012properties} and the appearance of a field-induced spin reorientation detected via magnetic torque \cite{PhysRevB.83.094425,vyaselev2017interplay}. However, as we elaborate in this work, the angle-dependence of the torque and specific pattern of NMR resonances are incompatible with conventional collinear magnetic orders.
The precise magnetic structure of the BETS at ambient pressure, and the role of Mn $S=5/2$ spins in the anion layer therefore remain open questions \cite{PhysRevB.83.094425,vyaselev2017interplay}. 

In order to address these questions, we first present {\it ab initio} calculations and specific heat measurements that point toward negligible coupling between the BETS and Mn spins. We then establish the minimal magnetic model for the BETS layers including all higher order couplings, and show that the classical ground state exhibits four-sublattice chiral order analogous to the spin-vortex crystal (SVC) observed in some Fe-based superconductors \cite{lorenzana2008competing,fernandes2016vestigial,meier2018hedgehog}.
Finally, we show that the $^{13}$C NMR \cite{vyaselev2012staggered,vyaselev2012properties} and magnetic torque \cite{PhysRevB.83.094425,vyaselev2017interplay} experiments are only compatible with this chiral order, thus confirming the ground state of $\kappa$-Mn.

{\it Role of Mn spins --} For $\kappa$-Mn, the magnetic couplings (pictured in Fig.~\ref{fig:structure}) can be divided into three categories:  $\pi$-$\pi$ (between BETS dimers), $d$-$d$ (between Mn ions),  and $\pi$-$d$ (between Mn and BETS). The latter two can be summarized as:
\begin{align}
\mathcal{H}^{\pi d} + \mathcal{H}^{dd} = \sum_{in} J_{ij}^{\pi d} \  \mathbf{s}_i\cdot\mathbf{S}_n+  \sum_{nm} J_{nm}^{dd} \  \mathbf{S}_n \cdot \mathbf{S}_m
\end{align}
where $\mathbf{s}_i$ is a BETS spin ($S=1/2$) at site $i$, $\mathbf{S}_n$ is a Mn spin ($S=5/2$) at site $n$. Following the approach of \cite{mori2002estimation}, we have estimated $J^{dd}$ and $J^{\pi d}$ using hopping parameters obtained from density functional theory calculations. Full details are given in \cite{sup}. Within the Mn layer, there are two dominant couplings (Fig.~\ref{fig:structure}(b)); we calculate $J_1^{dd} \approx J_2^{dd}\sim$ 0.5 - 1.0\,K. These interactions are both frustrated and disordered, due to random arrangements of the N(CN)$_2$ bridges. For the $\pi$-$d$ interactions, we also estimate a very small magnitude of $J_n^{\pi d}< 0.1$\,K, which suggests the BETS and Mn are essentially decoupled. The weak $\pi$-$d$ and $d$-$d$ couplings are consistent with weak antiferromagnetic tendencies of the Mn spins (experimentally, $\Theta_{\rm Mn} = - 2\langle S^2\rangle J^{dd} \sim-$5~K) \cite{PhysRevB.83.094425}. 
It is therefore expected that the Mn spins remain disordered until temperatures well below the ordering of the BETS spins.

\begin{figure}[b]
\includegraphics[width=\linewidth]{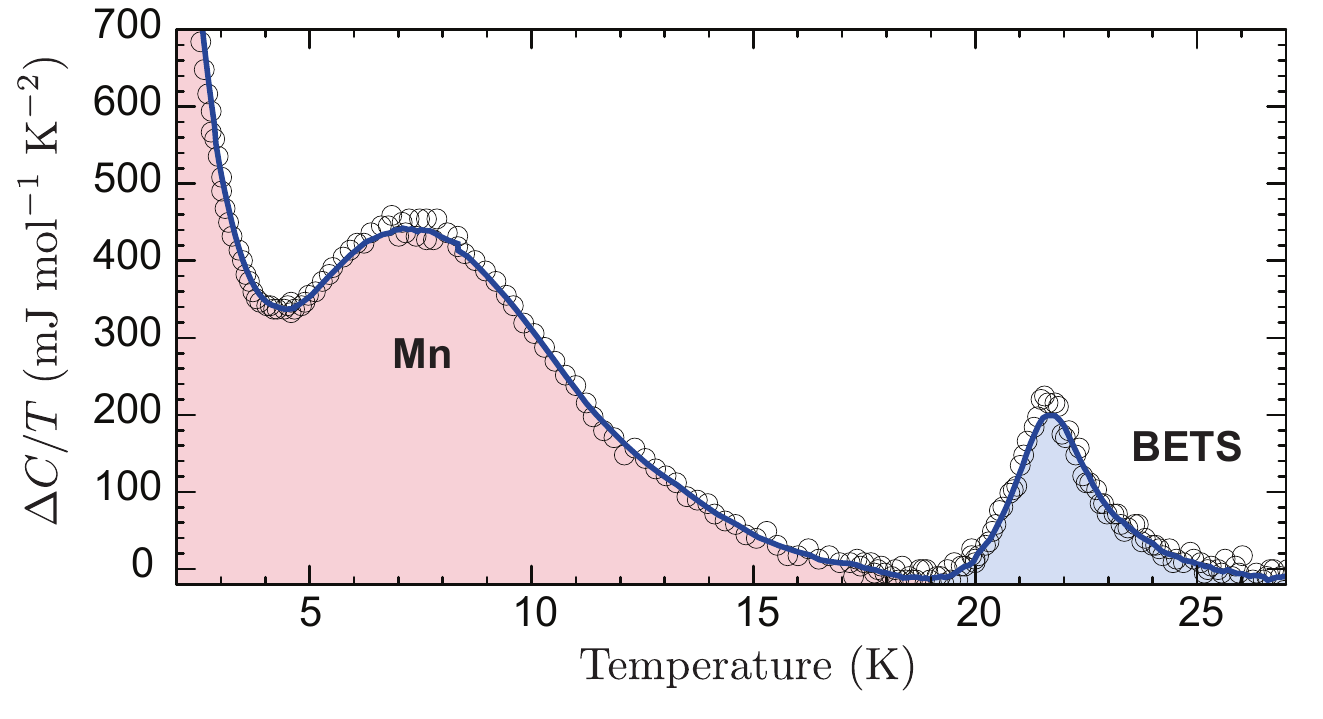}
\caption{Specific heat of $\kappa$-Mn. A smooth phononic background contribution was subtracted from the measured data to obtain $\Delta C$ (see \cite{sup}). The pink (blue) area indicate the estimate of entropy that is associated with the Mn (BETS) degrees of freedom.}
\label{fig:CV}
\end{figure}

In order to assess these energy scales and the coupling of the BETS and Mn spins experimentally, we measured the specific heat of $\kappa$-Mn. In Fig.~\ref{fig:CV}, we show an estimate of the electronic and magnetic specific heat divided by temperature, $\Delta C/T$. This was obtained by subtracting a smooth background function that serves as a proxy for phononic contributions (see \cite{sup} for details). The peak in $\Delta C/T$ at $T_{\rm N} \sim 21.5$ K signals the MIT that occurs concomitantly with the magnetic ordering of the BETS spins. The entropy change is estimated as $\Delta S_{\rm MIT} \approx \int_{19}^{25} \Delta C/T \ dT \approx (0.4\,\pm\,0.1)$ J$\cdot$mol$^{-1}\cdot$K$^{-1}$ (8\% of $R \ln 2)$, which is too small to indicate significant participation of the Mn spins. Instead, the entropy change is comparable to $\kappa$-ET salts with non-magnetic anions; for example, a change of electronic entropy of $\Delta S \approx 0.25$ J$\cdot$mol$^{-1}\cdot$K$^{-1}$ was measured across the (charge-order) MIT in an ET-based salt \cite{Gati18}, while magnetic ordering in the insulating $\kappa$-Cl was reported to have negligible $\Delta S$ \cite{yamashita2010heat}. 

For comparison, $\Delta S_{\rm MIT}$ is one order of magnitude smaller than the value observed for \mbox{$\lambda$-(BETS)$_2$FeX$_4$} (X = Cl, Br), where sizeable $\pi$-$d$ interactions lead to simultaneous ordering of the Fe$^{3+}$ and $\pi$ system \cite{mori2002estimation,konoike2004magnetic,kartsovnik2016interplay}. In these $\lambda$-phase materials, $\pi$-$d$ coupling also produces additional signatures that are absent in $\kappa$-Mn: (i) no field-induced Jaccarino-Peter superconductivity \cite{PhysRevB.82.155123,PhysRevB.83.094425}, and (ii) no beats in Shubnikov-de Haas effect  \cite{zverev2019fermi}. 
We therefore conclude that the $\pi$-$d$ coupling is sufficiently weak that the Mn spins play no significant role in the BETS magnetism.

Below $T_{\rm N}$, a separate broad feature in $\Delta C/T$ appears centered around 8 K, followed by a pronounced increase of $\Delta C/T$ below $\sim 4$ K. Given limitations in the lowest-accessible temperature, we can provide only a lower bound of the associated entropy: $\Delta S_{\rm Mn} > \int_{2}^{19} \Delta C/T \ dT \approx 4.4$ J$\cdot$mol$^{-1}\cdot$K$^{-1}$ (29\% of $R \ln 6$). This distinctly larger entropy change can only be associated with growing antiferromagnetic correlations between Mn spins. The temperature scales are compatible with the computed $J^{dd}$ couplings and measured $\Theta_{\rm Mn}$ \cite{PhysRevB.83.094425}. The multiple features in $\Delta C/T$ may reflect the combined frustration \cite{schmidt2015thermodynamics,Tutsch19} and disorder, with the increase below 4K potentially signifying the onset of freezing or ordering of the Mn spins.

{\it Magnetic Model for BETS --} Given the weak $\pi$-$d$ couplings, the magnetic order within the organic layer must arise from $\pi$-$\pi$ couplings:
\begin{align}\label{eqn:BETS}
\mathcal{H}^{\pi\pi} =& \  \sum_{ij} \left(J_{ij}^{\pi\pi} \ \mathbf{s}_i \cdot \mathbf{s}_j + \mathbf{D}_{ij}^{\pi\pi} \cdot \mathbf{s}_i \times \mathbf{s}_j + \mathbf{s}_i \cdot \mathbf{\Gamma}_{ij}^{\pi\pi} \cdot \mathbf{s}_j \right)  \nonumber \\
 & \ + \frac{1}{S^2}\sum_{ijkl} K_{ijkl}^{\pi\pi}  P_{ijkl} \\
 P_{ijkl} = & \ [(\mathbf{s}_i \cdot \mathbf{s}_j) (\mathbf{s}_k \cdot \mathbf{s}_l)+ (\mathbf{s}_j \cdot \mathbf{s}_k) (\mathbf{s}_i \cdot \mathbf{s}_l)   \nonumber \\
& \  \hspace{10mm}-(\mathbf{s}_i \cdot \mathbf{s}_k) (\mathbf{s}_j \cdot \mathbf{s}_l)  ]
\end{align}
where $K_{ijkl}^{\pi\pi}$, $\mathbf{D}_{ij}^{\pi \pi}$, $\mathbf{\Gamma}_{ij}^{\pi \pi}$, parameterize the 4-site ring exchange, Dzyaloshinskii–Moriya (DM) interaction, and pseudo dipolar coupling, respectively. The unique exchange constants are defined according to Fig.~\ref{fig:structure}(a). 

In order to estimate the magnitudes of the couplings, we employed the {\it ab initio} method outlined in \cite{winter2017importance}, (see \cite{sup} for full details). The results are summarized as follows. For the isotropic couplings, we estimate
$J=260\,$K, $J^\prime=530\,$K, $J^{\prime \prime}=4.7\,$K, $J^{\prime \prime \prime}=26\,$K, $K=16\,$K and $K^\prime=39\,$K. For the anisotropic couplings, the presence of a crystallographic inversion center ensures that $|\mathbf{D}^\prime| = 0$ and $|\Gamma^\prime| \approx 0$. As a result, the only significant anisotropic couplings appear for the nearest neighbor dimers, with $\mathbf{D} = (\pm 22.6,\mp 1.9,\pm 8.8)\,$K oriented approximately along the long axis of the dimers. The specific orientations of each $\mathbf{D}$ vector is indicated in Fig.~\ref{fig:supp_dms} of the Supplementary Information \cite{sup}. The largest entry in $\Gamma$ is 0.6\,K; the full tensors are given in \cite{sup}.  The larger magnitude of the anisotropic couplings compared to ET salts~\cite{winter2017importance} is due to enhanced spin-orbit coupling afforded by the heavy Se atoms in the BETS molecules.

\begin{figure}[b]
\includegraphics[width=\linewidth]{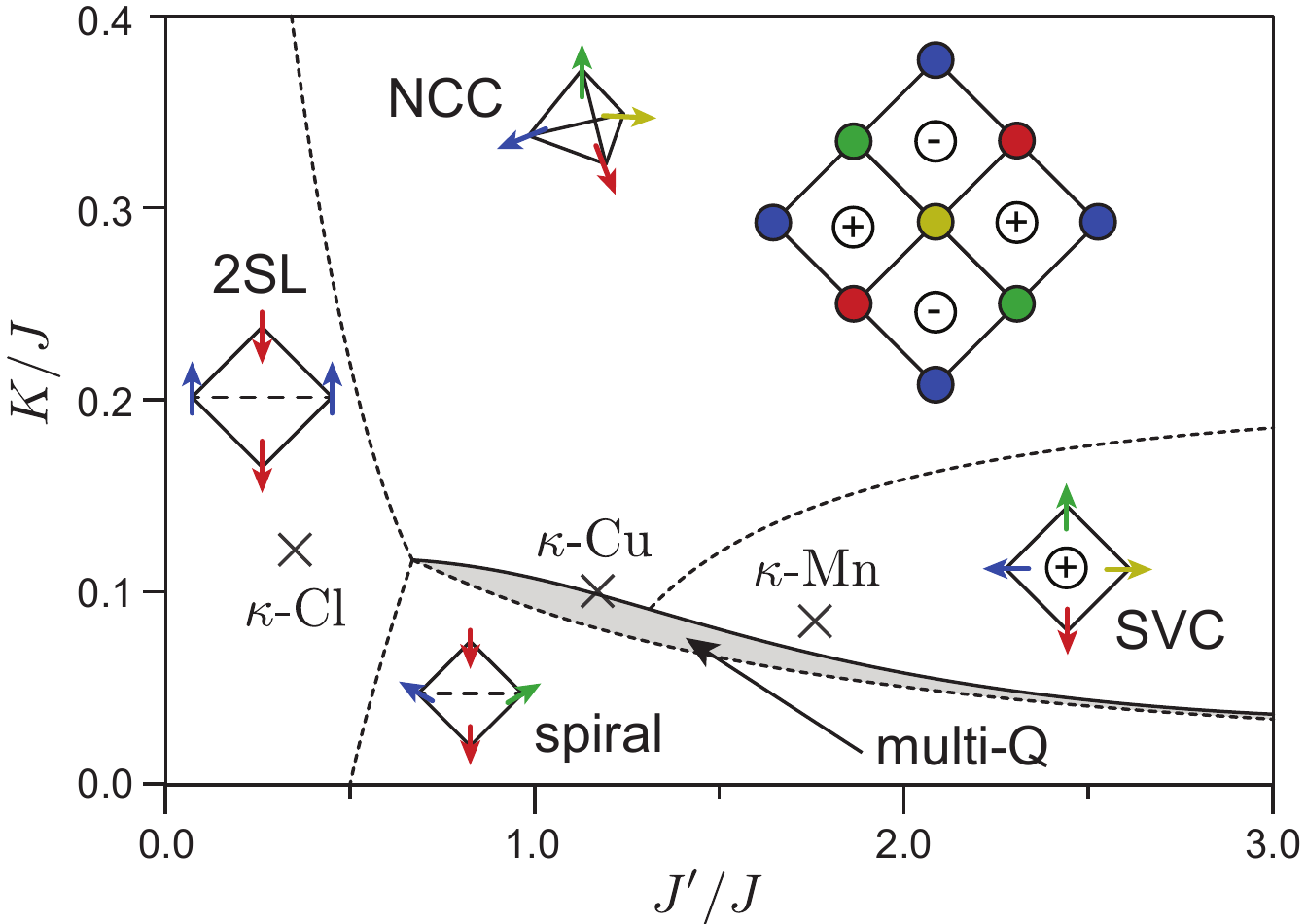}
\caption{Classical phase diagram for the model defined by Eq.~\ref{eqn:BETS}, with the constraints $J^\prime/J = J^{\prime\prime\prime}/J^{\prime\prime} = K^\prime/K$, and $K/J^{\prime\prime} = 2$, and $|\mathbf{D}| = |\mathbf{\Gamma}| = 0$. The approximate locations of various organic materials are indicated. 2SL = two-sublattice Neel order, SVC = four-sublattice coplanar spin-vortex crystal, NCC = four-sublattice non-coplanar chiral. Dashed (solid) lines indicate transitions expected to be continuous (first order). The staggered pattern of vector chirality $\mathbf{v}_p$ for the NCC and SVC phases is indicated by $\oplus,\ominus$ in the inset. The location of $\kappa$-Cu and $\kappa$-Cl are is based on \cite{winter2017importance,riedl2019critical}.
}
\label{fig:phase}
\end{figure}

Based on the computed couplings, we find that the classical ground state for $\kappa$-Mn is a four-sublattice configuration of orthogonal spins, which we label as spin-vortex crystal (SVC). To put this state into context, in Fig.~\ref{fig:phase}, we show the classical phase diagram of model (\ref{eqn:BETS}), taking the anisotropic couplings to zero, and using the approximate ratios of the isotropic couplings suggested from perturbation theory \cite{holt2014spin}. Various regions of the phase diagram have been studied previously \cite{roger1989cyclic,chubukov1992phase,kubo1997ground,kubo1998possible,kubo2003magnetic,lauchli2005two,holt2014spin,yasuda2018ground,messio2011lattice}.
The limit $J \gg J^\prime$ corresponds to the square lattice, for which the ground state is a collinear two-sublattice (2SL) N\'eel order. For small $K$, this is bordered by a family of coplanar spiral states, typically with incommensurate wavevectors. This family includes, as a special case for the triangular lattice ($J^\prime/J$=1), conventional 120$^\circ$ order. 
Starting from this point and increasing $K$ leads first to a narrow multi-Q state with a modulated canting of spins out of the plane of the spiral \cite{kubo2003magnetic}. 
For large $K$, there is a noncoplanar chiral (NCC) phase. The spin orientation in the NCC phase can be understood as follows: starting from the 2SL state with spins oriented perpendicular to the plane, each spin is then canted towards the plane to form four sublattices in the pattern indicated in Fig.~\ref{fig:phase}. For the special case $J^\prime/J$=1, spins on different sublattices satisfy $\mathbf{s}_i\cdot\mathbf{s}_j = -S^2/3$, as if oriented along the vertices of a tetrahedron
\cite{kubo1997ground,kubo1998possible,kubo2003magnetic}. 
 With increasing $J^\prime$, the spins completely tilt towards a common plane, leading to the coplanar spin-vortex crystal (SVC) \cite{lauchli2005two,chubukov1992phase}. $\kappa$-Mn differs from other organics primarily in terms of $J^\prime/J \approx (t^\prime/t)^2$, which can been estimated
 as $\sim 0.2 - 0.35$ for $\kappa$-Cl \cite{kandpal2009revision,nakamura2009abinitio,winter2017importance,jacko2020interplay}, $\sim 0.8 - 1.2$ for $\kappa$-Cu \cite{kandpal2009revision,nakamura2009abinitio,koretsune2014evaluating,winter2017importance}, and $\sim 2$ for $\kappa$-Mn.
 Recent DMRG studies \cite{szasz2020chiral,szasz2021phase} of the Hubbard model have found a similar phase diagram, enriched by QSL states.

{\it Order Parameter --} The SVC and NCC phases are distinguished from the other phases by a finite vector chiral order parameter. Specifically, for each 4-site square plaquette $p$ defined by the $J$-bonds (solid lines, Fig.~\ref{fig:structure}(a)), we define a vector chirality $\mathbf{v}_p = \mathbf{s}_1 \times \mathbf{s}_2 + \mathbf{s}_2 \times \mathbf{s}_3+ \mathbf{s}_3 \times \mathbf{s}_4 + \mathbf{s}_4 \times \mathbf{s}_1$. The SVC and NCC phases correspond to a staggered pattern of neighboring $\mathbf{v}_p$ vectors, depicted by $\oplus, \ominus$ in the inset of Fig.~\ref{fig:phase}. To see why large $K$ favors finite $|\mathbf{v}_p|$, it is useful to write, for a given plaquette:
\begin{align}
    |\mathbf{v}_p|^2 = \frac{3}{2}-\sum_{i\in\{1..4\}}  \frac{1}{2} \mathbf{s}_i \cdot \mathbf{s}_{i+1}-  \mathbf{s}_i \cdot \mathbf{s}_{i+2} \nonumber \nonumber\\   
    -2\left[ P_{1234}-(\mathbf{s}_1 \cdot \mathbf{s}_3)(\mathbf{s}_2 \cdot \mathbf{s}_4)\right] 
\end{align}
In general, the ring exchange $K$ and couplings $J^\prime, J^{\prime\prime\prime}$ are minimized when nearest neighbor spins are orthogonal $\langle \mathbf{s}_i \cdot \mathbf{s}_{i+1}\rangle \sim 0$, but second neighbor spins are antiparallel ($\langle \mathbf{s}_i \cdot \mathbf{s}_{i+2}\rangle > 0, \langle P_{ijkl}\rangle < 0$), which corresponds to a finite value of $|\mathbf{v}_p|$. For the 2SL and spiral phases we have $|\mathbf{v}_p| = 0$.

Due to the specific symmetries of the crystal, states with finite $|\mathbf{v}_p|$ can be distinguished based on their magnetic anisotropy.
 Importantly, the periodicity of the DM-vector component in the $ac$-plane and the component along the $b$-axis is different by symmetry (see \cite{sup}). The vector chirality couples linearly only to $\mathbf{D}_b$, which pins $\mathbf{v}_p || b$ at low fields. For the coplanar SVC phase, this confines the spins to lie in the $a^*c$-plane. In contrast, the symmetry of the 2SL and spiral phases are such that they couple only to $\mathbf{D}_{ac}$. At low fields, the energy is minimized for spins confined to the plane perpendicular to  $\mathbf{D}_{ac}$. This distinction may be probed by the angular dependence of the magnetic torque $\tau(H,\theta)$. 

{\it Magnetic Torque:} While detailed analysis of recently reported $\tau(H,\theta)$ \cite{vyaselev2017interplay,PhysRevB.83.094425} is complicated by a background contribution from the paramagnetic Mn, we can make general observations. For $T<T_{\rm N}$, step-like features appear in $\tau(H)$ due to a field-induced reorientation of spins within the BETS layer. While this has been discussed as a spin-flop transition \cite{vyaselev2017interplay,anderson1964statistical,keffer1973dynamics} of an easy-axis 2SL phase, the angular dependence is not directly compatible with this scenario. In particular, the torque on the BETS spins vanishes for fields oriented in the entire $a^*c$-plane. This should be considered anomalous for two reasons: (i) for a conventional spin-flop, $\tau(H)$ vanishes by symmetry only for fields along the easy-axis, rather than an entire plane, and (ii) for the 2SL (and spiral) phases, the $a^*c$-plane is not a special plane of symmetry. Taken together, these findings suggest the field couples to an order parameter of different symmetry. Similar effects have also been seen in the related mixed Co/Mn salt \cite{kushch2017new}.

\begin{figure}[t]
\includegraphics[width=0.9\linewidth]{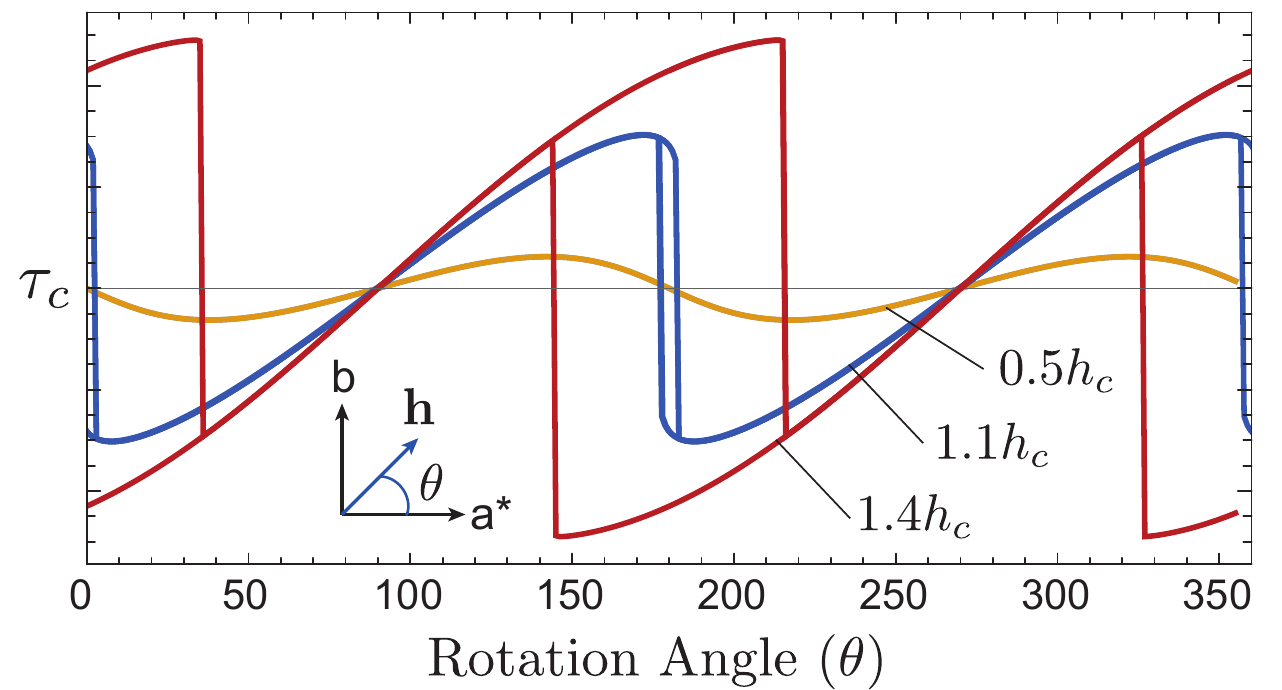}
\caption{Magnetic torque $dE_{\rm SVC}/d\theta$ as a function of rotation angle $\theta$ around the $c$-axis for $E_{\rm SVC}$ defined by Eq. (\ref{eqn:torque}). $h_c$ is the critical chirality-flop field at $\theta = 0$.}
\label{fig:torque}
\end{figure}

Instead, the observed torque can be easily explained by an ordered phase with finite $|\mathbf{v}_p|$. By symmetry (see \cite{sup}), the free energy can be written schematically as:
\begin{align}\label{eqn:torque}
E_{\rm SVC} = -(0,d_b,0)\cdot \mathbf{v} - (\mathbf{h}\cdot \mathbf{v})^2 + \mathcal{O}(h^4)
\end{align}
where $\mathbf{v}$ is the staggered vector-chirality, $d_b$ is a reduced DM-coupling, and $\mathbf{h}$ is a reduced external field. Due to the DM-interaction, $\mathbf{v}$ is pinned to the $b$-axis at low-fields. An applied field tends to tilt the ordering plane to be perpendicular to the field. For fields oriented close the $a^*c$-plane, $d_b$ competes with the field, leading to a rapid rotation of the ordering plane at a critical field $h_c$. Such a transition may be viewed as a flop of the vector-chirality. The associated torque ($\tau_c = dE_{\rm SVC}/d\theta$) for rotation around the $c$-axis is shown in Fig.~\ref{fig:torque}. For $h>h_c$ there are two metastable domains with $\mathbf{v}$ tilted towards and away from $\mathbf{h}$. This leads to widening hysteresis with increasing field, which may be related to the reported irreversibility of $\tau_c$ for field sweeps with $h>h_c$. At all fields, the average value of $\tau_c$ for $\mathbf{h}||a^*$ vanishes. Similarly, for rotation around the $b$-axis, the associated torque $\tau_b = 0$ for all field orientations.

\begin{figure}[t]
\includegraphics[width=\linewidth]{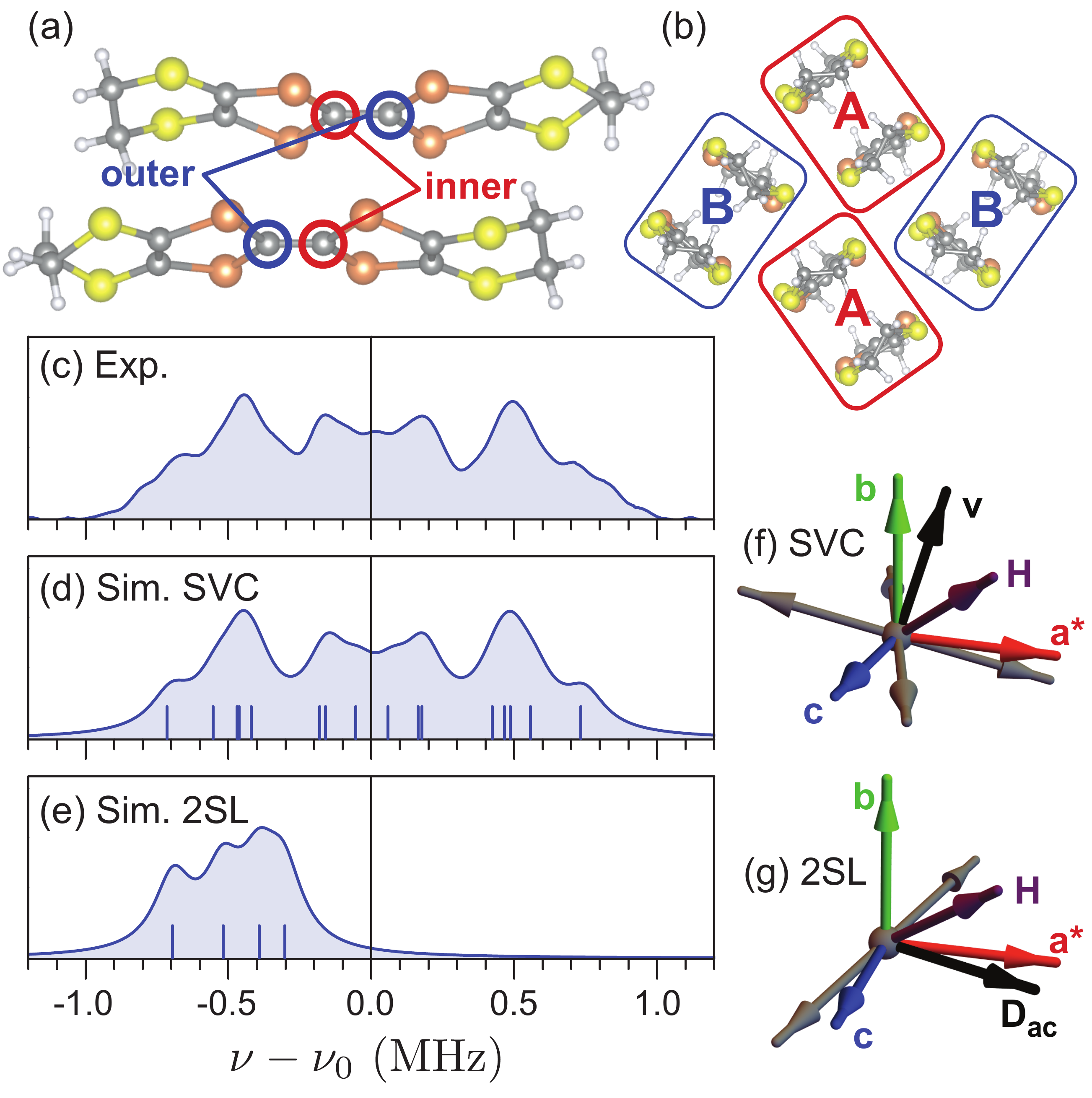}
\caption{(a) Distinct $^{13}$C sites in each BETS dimer. (b) Dimer sublattices. (c) Experimental $^{13}$C NMR spectra measured at 5 K and 7 T for field oriented 45$^\circ$ from the $a^*$ axis, and $\perp[0\bar{1}1]$ obtained from \cite{vyaselev2012properties}. (d,e): Simulated spectra for SVC, and 2SL phases (see text). Sticks indicate resonance positions. (f,g) Orientations of the sublattice moments (grey) for the simulations.}
\label{fig:nmr}
\end{figure}

{\it $^{13}$C NMR:} As a more selective probe of the magnetic order in the BETS layer, we also consider the $^{13}$C NMR data
reported in Refs.~\onlinecite{vyaselev2012staggered,vyaselev2012properties}. 
Below $T_{\rm N}$, the resonance frequency of each $^{13}$C nucleus $n$ at dimer site $i$ shifts by $\Delta \nu_{n,i}$ from the natural Larmor frequency due to the hyperfine coupling with the local spin moment:
\begin{align}
\Delta\nu_{n,i} = \gamma_C \left( | \mathbf{H}_{\text{eff},n,i} | - |\mathbf{H}_{\rm ext}|\right) \\
\mathbf{H}_{\text{eff},n,i} = \mathbf{H}_{\rm ext} + \mathbb{A}_{i,n} \cdot \langle \mathbf{s}_i\rangle
\end{align}
where $\gamma_C = 10.7084$ MHz/T is the gyromagnetic ratio, $\langle \mathbf{s}_i\rangle$ is the local spin expectation value, and $\mathbb{A}_{i,n}$ is the local hyperfine coupling tensor.
There are two types of isotopically enriched $^{13}$C sites per dimer, depicted in Fig.~\ref{fig:nmr}(a). In addition, as shown in Fig.~\ref{fig:nmr}(b), there are two symmetry-related dimers per unit cell (sublattice A and B). Together, this yields four crystallographically distinct $^{13}$C sites per unit cell. In order to analyze the reported spectra, we first estimated the $\mathbb{A}$ tensors using ORCA \cite{neese2020orca} (see \cite{sup} for details). We then simulated the expected resonance patterns for different magnetic configurations. For this purpose, we employ a Lorentzian broadening consistent with the experimental widths, and have ignored the Mn dipolar fields shown to be relevant for $^{1}H$ NMR. 
In Fig.~\ref{fig:nmr}(c), we show the experimental spectra for $T = 5$ K, and $H = 7$ T, reproduced from Ref.~\onlinecite{vyaselev2012properties}. The field is oriented $45^\circ$ degrees from the $a^*$-axis, and perpendicular to the $[0\bar{1}1]$ direction. It is well below the spin-reorientation transition at this angle. 
The resonance is symmetrical about the Larmor frequency, with a rich fine structure, indicating many magnetically inequivalent $^{13}$C sites.

From the symmetry of the resonances, the 2SL phase can be immediately ruled out. In $\kappa$-phase organics, the A and B dimers in the unit cell correspond precisely with the two magnetic sublattices in the 2SL state. At moderate fields, a single magnetic domain is selected by the DM-interaction $\mathbf{D}_{ac}$, which selects a unique preferred orientation of the sublattice moments with respect to the canted moment. As a result, the magnetically inequivalent $^{13}$C sites are in 1:1 correspondence with the crystallographically distinct sites. This leads to  four distinct resonances, with {\it asymmetrical} shifts for most field directions. This property was previously employed to confirm 2SL order in $\kappa$-Cl \cite{miyagawa2004nmr}. In Fig.~\ref{fig:nmr}(e) we show a representative spectrum for $\kappa$-Mn in the 2SL phase, assuming the sublattice moments are oriented along $\pm \mathbf{D}_{a^*c} \times \mathbf{H}$, as shown in Fig.~\ref{fig:nmr}(g). As can be seen, the simulated spectrum is completely incompatible with the experiment.

We next consider 4-sublattice SVC. In this state, there is not a unique correspondence between the crystallographic and magnetic sublattices; domains are expected in which the A and B sublattices are populated with spins of all four orientations. In total, this leads to 16 distinct resonances for general field orientations. The magnetic structure is symmetric under the combination of inversion and time-reversal, which ensures that the NMR resonances are symmetrically distributed. In order to evaluate whether the experimental spectra is compatible with SVC, we fit the data for a four magnetic sublattice model using the computed $\mathbb{A}$ tensors. To ensure a symmetrical spectrum, we only constrained $\mathbf{s}_1 = -\mathbf{s}_2$ and $\mathbf{s}_3 = -\mathbf{s}_4$. The resulting best fit, shown in Fig.~\ref{fig:nmr}(d) shows almost perfect agreement with the experiment. The fitted ordered moment is only $\langle\mathbf{s}_i\rangle = 0.15\  \mu_B$, which suggests strong quantum and/or thermal fluctuations.
More importantly, the orientations of the fitted moments (Fig.~\ref{fig:nmr}(f)) conform with the expectations for SVC. The vector chirality $\mathbf{v}$ is oriented close to the $b$-axis, but tilted towards the external field $\mathbf{H}$. The fitted moments form an angle $\angle \mathbf{m}_1\mathbf{m}_3 = 84^\circ$, which is close to the ideal of $90^\circ$. On this basis, we conclude that SVC represents the best fit of the reported NMR spectra.

{\it Discussion:} On balance, we find that the anomalous $^{13}$C NMR, and magnetic torque response are both consistent with SVC order in $\kappa$-Mn. This is compatible with {\it ab initio} results; consideration of higher order 4-site ring exchange places $\kappa$-Mn in a region of the classical phase diagram displaying spin vortex crystal order. Furthermore, we have ruled out the possibility that these features arise from coupling of the BETS $S=1/2$ spins with the Mn $S=5/2$ spins in the anion layer. Both theoretical estimates of the $\pi$-$d$ magnetic couplings, and the temperature dependence of the magnetic specific heat, indicate exceedingly weak coupling between the two subsystems. 

All together, these observations serve as a definitive proof of the importance of ring-exchange in the $\kappa$-phase organic materials.
While below the infamous 6\,K anomaly of $\kappa$-Cu, a description in terms of a valence bond solid phase has been put forward~\cite{riedl2019critical,miksch2021gapped}, the recent proposals \cite{szasz2020chiral,chen2021quantum,szasz2021phase,wietek2021mott,cookmeyer2021four} of a (gapped) CSL state should be considered seriously above 6\,K, given that the parent chiral ordered state has now been observed in \mbox{$\kappa$-Mn}.

{\it Acknowledgements $-$ }  We acknowledge useful discussions with Y. Agarmani. S.M.W. acknowledges support through an NSERC Canada
Postdoctoral Fellowship. K.R., M.L. and R.V. acknowledge support by the Deutsche Forschungsgemeinschaft (DFG, German Research Foundation) for funding through TRR 288 --- 422213477 (projects A05, A06). O.M.V., N.D.K., and M.V.K. acknowledge financial support from
the German Research Foundation (Deutsche Forschungsgemeinschaft, DFG) via grant KA 1652/5-1 and from the Russian Foundation for Basic Research, grant No. 21-52-12O27. N.D.K. also acknowledges the support of the State Assignment of the topic No. AAAA-A19-11902390079-8.

\bibliography{BETS}

\clearpage

\appendix
\section{Supplemental Information}
\subsection{Magnetic Couplings}
The magnetic Hamiltonian can be described by three types of couplings: $d$-$d$ (between Mn ions),  $\pi$-$\pi$ (between BETS dimers), and $\pi$-$d$ (between Mn and BETS).
\begin{align}
\mathcal{H} =& \  \sum_{ij} \left(J_{ij}^{\pi\pi} \ \mathbf{s}_i \cdot \mathbf{s}_j + \mathbf{D}_{ij}^{\pi\pi} \cdot \mathbf{s}_i \times \mathbf{s}_j + \mathbf{s}_i \cdot \mathbf{\Gamma}_{ij}^{\pi\pi} \cdot \mathbf{s}_j \right)  \nonumber \\
 & \ +\sum_{in} J_{in}^{\pi d} \  \mathbf{s}_i\cdot\mathbf{S}_n+  \sum_{nm} J_{nm}^{dd} \  \mathbf{S}_n \cdot \mathbf{S}_m \nonumber\\
 & \ + 4\sum_{ijkl} K_{ijkl}^{\pi\pi} [(\mathbf{s}_i \cdot \mathbf{s}_j) (\mathbf{s}_k \cdot \mathbf{s}_l) \nonumber \\
& \  \hspace{10mm}+ (\mathbf{s}_j \cdot \mathbf{s}_k) (\mathbf{s}_i \cdot \mathbf{s}_l)  -(\mathbf{s}_i \cdot \mathbf{s}_k) (\mathbf{s}_j \cdot \mathbf{s}_l)  ]
\end{align}
where $\mathbf{s}_i$ is a BETS spin ($S=1/2$) at site $i$, $\mathbf{S}_n$ is a Mn spin ($S=5/2$) at site $n$. In order to estimate the couplings, we first computed hopping integrals employing two methods: (i) for $d$-$d$ and $d$-$\pi$ hoppings, we employed the full potential local orbital (FPLO) basis~\cite{fplo}, generalized gradient approximation (GGA) exchange correlation functional~\cite{Perdew1996} and projective Wannier functions~\cite{Eschrig2009}; (ii) for the $\pi$-$\pi$ hoppings we employed ORCA \cite{neese2020orca} calculations on dimer pairs at the B3LYP/def2-SVP level, using the scheme outlined in \cite{winter2017importance}. This latter method allows for the incorporation of spin-orbit coupling (SOC) required to address the anisotropic couplings $\mathbf{D}_{ij}^{\pi\pi}$ and $\mathbf{\Gamma}_{ij}^{\pi\pi}$. All calculations were based on the room temperature structure reported in \cite{kushch2008pi}. In this structure, there is disorder in both the positions of the N(CN)$_2$ ligands and ethylene endgroups of the BETS molecules. For the ligands, one out of three dicyanoamine chains has 50\% occupancy on each of two positions which are equivalent within $P2_1c$ space group; lowering the space group to $P2_1$ or to $P_c$ is necessary to achive full occupancy; we choose the former.  For the ethylene end groups of which every other is disordered, we perform the calculations for the majority (80\%) configuration. 

We first discuss the $d$-$d$ couplings. The Mn atoms form a distorted triangular lattice bridged by dicyanamide ions, with two distinct nearest neighbor bonds having Mn-Mn distances of 7.370 \AA \ and 8.412 \AA. The hoppings between Mn $d$-orbitals are presented in Table \ref{tab_Mnhop}. Following Ref.~\onlinecite{mori2002estimation}, the magnetic couplings can be estimated using:
\begin{align}
J_{dd}=\frac{4}{25U_d}\sum_{i=1}^{25} t_i^2
\end{align}
where $U_d$ is the average Coulomb repulsion in the $d$-orbitals. Employing $U_d \sim 3-6$ eV, we arrive at:
\begin{align}
J_{dd}(1) = +0.76 \text{ K} \text{ to } +1.5 \text{ K} \\ J_{dd}(2) = +0.59 \text{ K} \text{ to } +1.2 \text{ K}
\end{align}
These $d$-$d$ couplings are both small and geometrically frustrated, which is consistent with the lack of magnetic order in the Mn lattice at measured temperatures. 

We next consider the $\pi$-$d$ couplings. The relevant hoppings are shown in Table \ref{tab_MnBETShop}. Following Ref.~\onlinecite{mori2002estimation}, the magnetic couplings can be estimated using:
\begin{align}
J_{\pi d} = \frac{4}{5\Delta_{\pi d}} \sum_{i=1}^{5} t_i^2
\end{align}
were $\Delta_{\pi d}$ is the charge transfer energy between the BETS and Mn. Here we approximate $\Delta_{\pi d} \approx U_d$, yielding:
\begin{align}
J_{\pi d}(1) = +0.04\text{ K}\text{ to } +0.08 \text{ K}\\ J_{\pi d}(2) = +0.02\text{ K}\text{ to } +0.04 \text{ K} \\ J_{\pi d}(3) =   +0.02\text{ K} \text{ to } +0.04 \text{ K}
\end{align}
As a result of the large separation of Mn and BETS (and consequently small hopping integrals), the $\pi$-$d$ couplings are essentially negligible. These results support the experimental findings that the two subsystems are essentially decoupled. The computed couplings are orders of magnitude smaller than those estimated for \mbox{$\lambda$-(BETS)$_2$FeX$_4$} in Ref.~\onlinecite{mori2002estimation} following the same method.

\begin{table}[t]
\caption {\label{tab_Mnhop}Mn-Mn Hopping parameters (meV) obtained from FPLO.}
\centering\def\arraystretch{1.0}\small
\begin{ruledtabular}
\begin{tabular}{llrrrrrr}
Mn$-$Mn (1):&  & $d_{xy}$& $d_{xz}$  & $d_{yz}$ & $d_{x^2-y^2}$&$d_{z^2}$  \\
(7.370 \AA)& $d_{xy}$& +2.4&+2.1 &+15.2 &$\sim 0$ & $\sim 0$ \\
&  $d_{xz}$&+44.2 & +0.5& +2.6& $\sim 0$&  $\sim 0$   \\
& $d_{yz}$& -1.6& +0.4& +0.2& $\sim 0$&  $\sim 0$ \\
& $d_{x^2-y^2}$& $\sim 0$& $\sim 0$& $\sim 0$& +11.7& +8.3   \\
& $d_{z^2}$& $\sim 0$& $\sim 0$& $\sim 0$& -6.0&-4.0   \\
\hline
Mn$-$Mn (2):&  & $d_{xy}$& $d_{xz}$  & $d_{yz}$ & $d_{x^2-y^2}$&$d_{z^2}$  \\
(8.412 \AA)& $d_{xy}$&+0.6 & +2.2&+1.8 &$\sim 0$ & $\sim 0$ \\
&  $d_{xz}$&-2.9 &-21.0 &-0.4 &$\sim 0$ & $\sim 0$    \\
& $d_{yz}$& -0.1&-13.6 & -15.0&$\sim 0$ & $\sim 0$  \\
& $d_{x^2-y^2}$& $\sim 0$& $\sim 0$& $\sim 0$&-0.9 & -3.7    \\
& $d_{z^2}$& $\sim 0$& $\sim 0$& $\sim 0$& -3.5& -31.7   \\

\end{tabular}
\end{ruledtabular}
\end{table}

\begin{table}[t]
\caption {\label{tab_MnBETShop} Mn-BETS hopping paramaters (meV) obtained from FPLO.}
\centering\def\arraystretch{1.0}\small
\begin{ruledtabular}
\begin{tabular}{llrrrrrr}
BETS$-$Mn (1):&  & $d_{xy}$& $d_{xz}$  & $d_{yz}$ & $d_{x^2-y^2}$&$d_{z^2}$  \\
(9.751 \AA)& $\pi$ MO& -2.3&+1.3& +3.8& &  \\
\hline
BETS$-$Mn (2):&  & $d_{xy}$& $d_{xz}$  & $d_{yz}$ & $d_{x^2-y^2}$&$d_{z^2}$  \\
(10.288 \AA)& $\pi$ MO&+1.3&-3.3& -0.2& &  \\
\hline
BETS$-$Mn (3):&  & $d_{xy}$& $d_{xz}$  & $d_{yz}$ & $d_{x^2-y^2}$&$d_{z^2}$  \\
(10.731 \AA)& $\pi$ MO&+2.6&+2.0& -0.3& &  
\end{tabular}
\end{ruledtabular}
\end{table}

\begin{figure}[t]
\includegraphics[width=\linewidth]{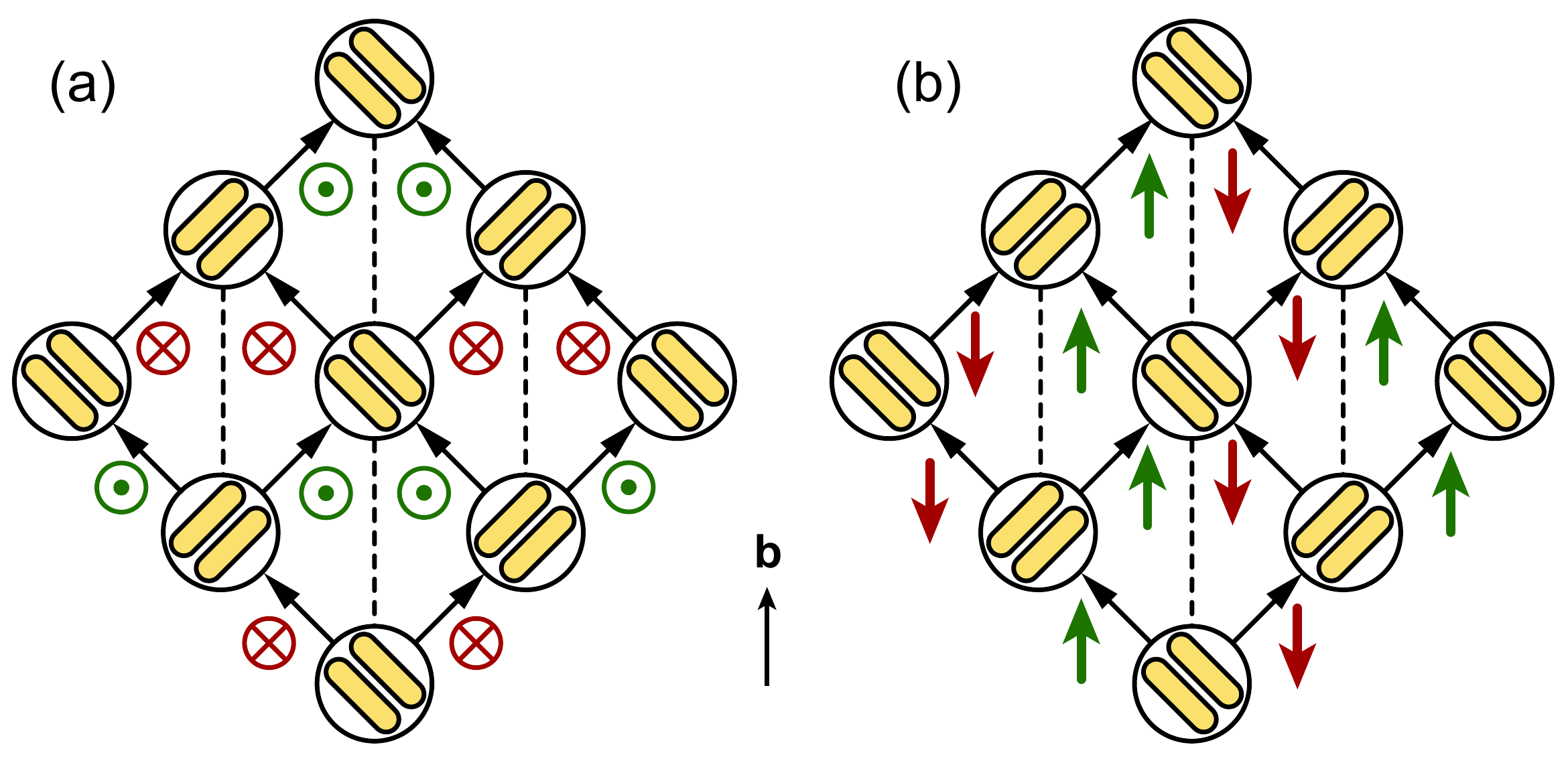}
\caption{Orientation of the DM-vectors for each bond, showing (a) the $a^*c$-component, and (b) the $b$-axis component. For each bond, the interaction is defined as $\mathbf{D}_{ij}\cdot (\mathbf{s}_i \times \mathbf{s}_j)$, with the black arrow pointing from site $i$ to site $j$. The $b$-axis component has same periodicity as a staggered vector-chirality.}
\label{fig:supp_dms}
\end{figure}

To estimate the couplings within the BETS layers, we followed the approach of \cite{winter2017importance}. In particular, spin-dependent hoppings were computed using ORCA and employed in linked cluster expansion (exact diagonalization) calculations on clusters of up to 4 dimers. For each cluster, the couplings of the low-energy spin Hamiltonian were extracted via projection. 
For the Coulomb couplings we considered a rescaled version of results from cRPA plus MLWO calculations for ET systems~\cite{nakamura2012abinitio}, successfully applied in Ref.~\onlinecite{winter2017importance} for various ET compounds. The parameter set consists of an on-site Hubbard repulsion $U = 0.55\,$eV, on-site Hund's coupling $J_{\rm H} = 0.2\,$eV, and nearest neighbour Hubbard repulsion $V= 0.15\,$eV. 
The bilinear $\pi$-$\pi$ magnetic couplings are given in Table~\ref{tab_PiPiCoupling} with the generalized bilinear exchange matrix in $\mathcal{H}=\sum_{ij} \mathbf{s}_i \cdot \Lambda_{ij} \cdot \mathbf{s}_j$ defined as:
\begin{align}
    \Lambda_{ij} =  
    \begin{pmatrix}
    J_{ij} + \Gamma_{ij}^{xx} & \Gamma_{ij}^{xy}+D_{ij}^z & \Gamma_{ij}^{xz}-D_{ij}^y \\
    \Gamma_{ij}^{xy}-D_{ij}^z & J_{ij} + \Gamma_{ij}^{yy} & \Gamma_{ij}^{yz}+D_{ij}^x \\
    \Gamma_{ij}^{xz}+D_{ij}^y & \Gamma_{ij}^{yz}-D_{ij}^x & J_{ij} - \Gamma_{ij}^{xx} - \Gamma_{ij}^{yy}
    \end{pmatrix}.
\end{align}
The specific orientation of the DM-vectors are depicted in Fig.~\ref{fig:supp_dms}. For a given 4-site plaquette formed by the $J$-bonds, the DM components in the $a^*c$ direction alternate for each bond, i.e. the total interaction can be written $D_{a^*c} \,(\pm \mathbf{s}_1 \times \mathbf{s}_2 \mp \mathbf{s}_2 \times \mathbf{s}_3\pm \mathbf{s}_3 \times \mathbf{s}_4 \mp \mathbf{s}_4 \times \mathbf{s}_1)_{a^*c}$. These components lead to a canting in the two-sublattice N\'eel phase. In the SVC and NCC phase these terms have zero expectation value. In contrast, the $b$-component takes the same sign for each bond, i.e. $\pm D_b \,(\mathbf{s}_1 \times \mathbf{s}_2 + \mathbf{s}_2 \times \mathbf{s}_3+ \mathbf{s}_3 \times \mathbf{s}_4 + \mathbf{s}_4 \times \mathbf{s}_1)_b=\pm D_b \,[\mathbf{v}_p]_b$. The free energy in the SVC and NCC phase is therefore restricted by symmetry to contain a linear coupling $(0,d_b,0)\cdot \mathbf{v}$, as presented in Eq.~(5) of the main text. Since the order parameter $\mathbf{v}$ is quadratic in spin, the lowest order of coupling to the magnetic field that respects time reversal symmetry is $(\mathbf{h}\cdot \mathbf{v})^2$.

\subsection{Comparison between ORCA and FPLO}
In previous works \cite{winter2017importance,riedl2019critical}, the application of ORCA to compute the hopping parameters including SOC for ET-salts proved to yield reliable magnetic couplings - particularly the magnitude and orientation of the DM-vector. However, for completeness, we also compare here the ORCA results with isotropic exchange parameters based on the non-relativistic hopping parameters obtained with FPLO. It should be noted that there are several differences in these two approaches: FPLO employs a GGA functional, and includes the full crystalline environment of each molecule, with Wannier functions constructed by projection onto an approximate linear combination of atomic orbitals. In contrast, with ORCA, we have employed a hybrid functional, with Wannier functions constructed via projection onto the precise molecular orbitals of isolated BETS molecules. The latter approach estimates pairwise hoppings from separate calculations on pairs of molecules, and therefore does not account for the full crystalline environment. Despite these major differences, the resulting hoppings are quite similar: from FPLO we estimate: $t_1 = 177$, $t_2 = 8$, $t_3 = 125$, $t_4 = 63$ meV, while ORCA produces: $t_1 = 236$, $t_2 = 19$, $t_3 = 153$, $t_4 = 64$ meV. The hoppings are numbered according to the convention in \cite{guterding2015influence}. The two approaches therefore yield almost the same ratio of $t^\prime /t \sim 0.57$ for FPLO and 0.54 for ORCA. However, all of the hopping integrals from FPLO are smaller by approximately 15 - 20\%. To compensate this discrepancy, it is necessary to rescale the two-particle parameters (which had been previously optimized \cite{winter2017importance} for use in combination with ORCA hoppings) to $U = 0.4$, $J_H = 0.15$, and $V = 0.1$ eV. Then, employing the FPLO hoppings, we estimate $J=263\,$K, $J^\prime=402\,$K, $J^{\prime \prime}=7.4\,$K, $J^{\prime \prime \prime}=33\,$K, $K=19\,$K and $K^\prime=43\,$K. 

In comparison to the results based on ORCA hoppings, the exchange values are rather similar, with a reduced $J^\prime/J$ ratio and an increased ring-exchange weight $K/J$ and $K^\prime/J$. Considering the classical state energies given below, these parameters place $\kappa$-Mn in the NCC phase, with a relatively small out-of-plane tilting angle $\theta=15^\circ$. This phase still has a large staggered vector chirality with $\mathbf{v} || b$, and thus would respond similarly in the magnetic torque to the SVC phase. In the $^{13}$C NMR, we would expect additional peaks to appear due to the reduction of symmetry, but these may be buried within the experimental linewidth. Therefore, although we find no specific evidence for finite out-of-plane moments, their possibility should not be completely ruled out. 

\begin{table}[t]
\caption {\label{tab_PiPiCoupling} $\pi$-$\pi$ magnetic couplings (K) between BETS dimers with respect to $(a,b,c^\ast)$ coordinates.}
\centering\def\arraystretch{1.0}\small
\begin{ruledtabular}
\begin{tabular}{lccc}
$\mathcal{J}$:&  $J$& $(D_x,D_y,D_z)$  & $(\Gamma_{xx},\Gamma_{xy},\Gamma_{xz},\Gamma_{yy},\Gamma_{yz})$   \\
&  +260&(+22.6,-1.9,+8.8)& (+0.6,-0.1,+0.4,-0.4,0.0)   \\
\hline
$\mathcal{J}^\prime$:&  $J$& $(D_x,D_y,D_z)$  & $(\Gamma_{xx},\Gamma_{xy},\Gamma_{xz},\Gamma_{yy},\Gamma_{yz})$ \\
& +531&-& -  \\
\hline
$\mathcal{J}^{\prime \prime}$:&  $J$& $(D_x,D_y,D_z)$  & $(\Gamma_{xx},\Gamma_{xy},\Gamma_{xz},\Gamma_{yy},\Gamma_{yz})$ \\
& +4.7&-& -  \\ 
\hline
$\mathcal{J}^{\prime \prime \prime}$:&  $J$& $(D_x,D_y,D_z)$  & $(\Gamma_{xx},\Gamma_{xy},\Gamma_{xz},\Gamma_{yy},\Gamma_{yz})$ \\
& +25.6&(+2.3,-0.3,+0.9)& (0.1,0.0,0.0,0.0,0.0)  
\end{tabular}
\end{ruledtabular}
\end{table}

\section{Classical State Energies}

\begin{figure}[t]
\includegraphics[width=\linewidth]{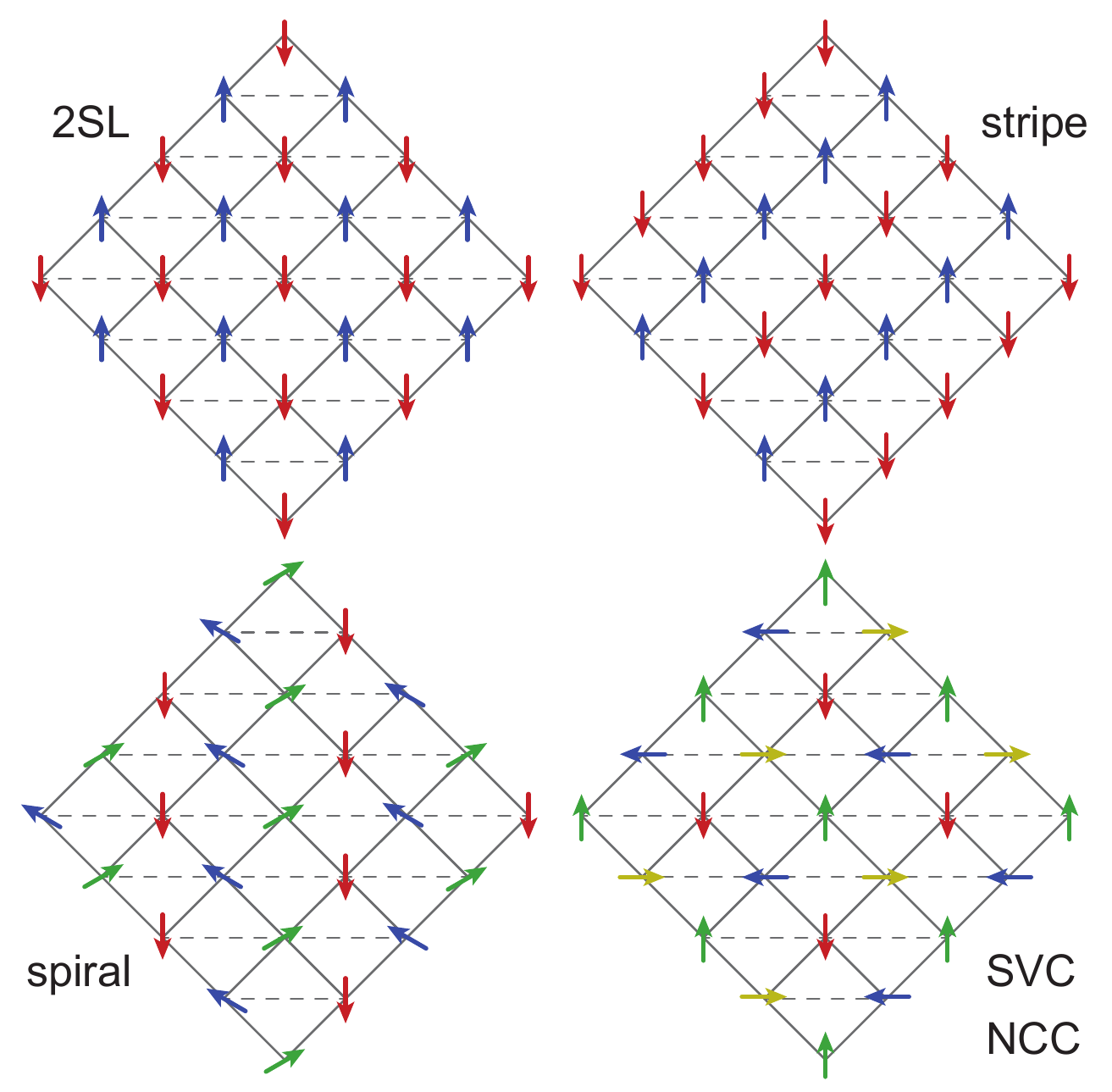}
\caption{Ordering patterns for the classical ground states mentioned in the text.}
\label{fig:supp_phase}
\end{figure}

In this section, we give the analytical expressions for the ground state energies (per site) for the classical states discussed in the main text (valid for antiferromagnetic signs of all $J$ couplings). These states are depicted in Fig.~\ref{fig:supp_phase}. 

First, we consider the orders identified in \cite{holt2014spin}. At small $K$, the ground states are the colinear two-sublattice ``$(\pi,\pi)$'' N\'eel order, and ``$(q,q)$'' spiral. The energies as a function of $q$ can be summarized by:
\begin{align}
\frac{E}{S^2} = & \ 2 J \cos \left(q \right)+ J^\prime \cos (2q) +J^{\prime\prime} + 2 J^{\prime\prime\prime} \cos\left(3q \right)\nonumber \\  & \  + K + 2 K^{\prime} 
\end{align}
where the ordering wavevector is given in the Brillouin zone of the primitive cell of the square lattice. Within the N\'eel phase, the minimum energy is obtained for $q = \pi$, yielding:
\begin{align}
    \frac{E_{\rm 2SL}}{S^2} =& \  -2 J + J^\prime + J^{\prime\prime} - 2 J^{\prime\prime\prime}  + K + 2 K^\prime
\end{align}
Within the spiral phase, the minimum energy $q$-vector is given by:
\begin{align}
q = \cos^{-1}\left(\frac{J_c-J^\prime }{12J^{\prime\prime\prime}} \right)
\end{align}
  where:
  \begin{align}
    J_c=& \ \sqrt{(J^\prime)^2 - 12 J J^{\prime\prime\prime} + 36 (J^{\prime\prime\prime})^2}
\end{align}
which yields:
\begin{align}
  \frac{E_{\rm sp}}{S^2} =& \ J^{\prime\prime}-\frac{J^\prime}{2} + K + 2 K^\prime+  \frac{ (J^\prime)^3 - 18 JJ^\prime J^{\prime\prime\prime} - J_c^3}{108 (J^{\prime\prime\prime})^2}
  \end{align}
At large $K$, there are three competitive states. The first is the two-sublattice collinear ``$(\pi,0)$'' stripe order, with energy:
\begin{align}
  \frac{E_{(\pi,0)}}{S^2}  = - J^\prime -  J^{\prime\prime} +  K + 2 K^\prime;
\end{align}
The chiral states have energy:
\begin{align}
\frac{E}{S^2} = & \ - J - J^{\prime\prime\prime} + \frac{K}{4} + K^\prime \nonumber \\ & \ 
+( J -  J^\prime +  J^{\prime\prime\prime} -  J^{\prime\prime} -  K) \cos (2 \theta) \nonumber \\ & \ + \left( K^\prime - \frac{K}{4}\right) \cos(4 \theta)
\end{align}
where $0 \leq \theta \leq \frac{\pi}{2}$ is the angle of tilting of the spins out of the plane. The N\'eel state is recovered by taking $\theta = \frac{\pi}{2}$. In the coplanar vector chiral spin-vortex crystal (SVC), $\theta = 0$. The energy is:
\begin{align}
    \frac{E_{\rm SVC}}{S^2} = -J^\prime - J^{\prime\prime} - K + 2 K^\prime
\end{align}
The SVC may be viewed as a multi-$q$ order that is a linear combination of $(\pi,0)$ and $(0,\pi)$ stripes. 
Classically, the SVC state is strictly lower in energy than the $(\pi,0)$ single-stripe phase for $K>0$. As a result, we do not find any region where the single-stripe phase represents the classical ground state. 
Finally, the four-sublattice non-coplanar chiral (NCC) order corresponds to the region of intermediate $\theta$ values. Assuming $\sign(J_a-J_b)=\sign(K-4K^\prime)$, the tilting angle can be expressed as:
\begin{align}
2\theta = \cos^{-1}\left(\frac{J_a-J_b}{K-4K^\prime} \right)
\end{align}
where:
\begin{align}
    J_a = & \ J+J^{\prime\prime\prime}\\
    J_b = & \ J^\prime + J^{\prime\prime}
\end{align}
This yields:
\begin{align}
    \frac{E_{\rm NCC}}{S^2} =& \  \frac{
 \frac{1}{2}(J_a - J_b)^2+ K J_b+ (K-2 J_a ) ( K - 2 K^\prime)}{ K - 4 K^\prime} 
 \end{align}

\subsection{$^{13}$C NMR Hyperfine Tensors}
In order to analyse the $^{13}$C NMR results, we first estimated the hyperfine coupling tensors for isolated [BETS$_2$]$^{1+}$ dimers via {\it ab initio} calculations using ORCA \cite{neese2020orca} at the B3LYP/EPR-II/def2-SVP level. Each dimer contains four $^{13}$C sites, with pairs of sites being related by inversion symmetry. As a result, there are two unique sites per dimer, conventionally labelled ``inner'' and ``outer'' (see \cite{miyagawa2004nmr}). Further, there are two dimers per unit cell (sublattice A and B), related by $2_1$ screw axis. The computed hyperfine tensors (units  kOe/$\mu_B$), in the crystallographic $(a^*,b,c)$ coordinate system are:
\begin{align}
\text{Dimer A, Outer:} \ \ \mathbb{A}_{\rm out}  = \left( \begin{array}{ccc}+0.9&+2.6&-1.6\\+2.6&+8.8&-6.4\\-1.6&-6.4&+4.1 \end{array}\right) \\ 
\text{Dimer A, Inner: } \ \ \mathbb{A}_{\rm in}  = \left( \begin{array}{ccc}-2.2&+0.8&-0.4\\+0.8&+2.0&-4.2\\-0.4&-4.2&+0.4\end{array}\right) \\
\text{Dimer B, Outer: } \ \ \mathbb{A}_{\rm out}  =\left( \begin{array}{ccc}+0.9&-2.6&-1.6\\-2.6&+8.8&+6.4\\-1.6&+6.4&+4.1 \end{array}\right) \\ 
\text{Dimer B, Inner: } \ \ \mathbb{A}_{\rm in}  =\left( \begin{array}{ccc}-2.2&-0.8&-0.4\\-0.8&+2.0&+4.2\\-0.4&+4.2&+0.4\end{array}\right)
\end{align}
The computed tensors may be validated by comparison with experimentally derived values for various ET salts \cite{mayaffre199413c,de1995c,miyagawa2004nmr,saito2016determination}. To facilitate this comparison, we rotate the computed tensors into the molecular coordinates of Ref.~\onlinecite{miyagawa2004nmr,saito2016determination}: the $x_m$-axis is taken to be parallel to the central C-C bond, and the $y_m$-axis is the perpendicular direction within the molecular plane. In these coordinates, we find:
\begin{align}
\mathbb{A}_{\rm out} = \left( \begin{array}{ccc}0.2&0.0&-0.6\\0.0&-0.4&-0.1\\-0.6&-0.1&+15.1 \end{array}\right) \\ 
\mathbb{A}_{\rm in} =\left( \begin{array}{ccc}-2.3&+0.0&+0.9\\+0.0&-3.3&+0.3\\+0.9&+0.3&+6.0\end{array}\right)
\end{align}
Not surprisingly, the hyperfine tensor is dominated by the $[\mathbb{A}]_{zz}$ component, as the unpaired electrons occupy $\pi$-orbitals with $p_z$ character in the molecular coordinate system \cite{miyagawa2004nmr}. Further, we find that $[\mathbb{A}_{\rm out}]_{zz} > [\mathbb{A}_{\rm in}]_{zz}$, consistent with the experimental trends for ET salts. The absolute magnitudes of the principle components are also consistent with those reported in \cite{mayaffre199413c,de1995c}. We therefore conclude that the estimated hyperfine tensors are of sufficient accuracy to simulate the experimental NMR spectra.

\subsection{Specific Heat Measurements}

\textit{Experimental Details - } Measurements of specific heat were performed by employing a high-resolution ac-modulation technique \cite{Sullivan68} on a single crystal of mass $m =$
(40\,$\pm$\,20)\,$\mu$g. Details of the setup, specially designed
for measuring very small plate-like crystals, such as $\kappa$-Mn, are presented in \cite{Mueller02}. Measurements were performed upon warming in the temperature range 1.8\,K$\,\leq\,T\,\leq\,$29\,K. For the measurements, the temperature oscillation amplitude $\Delta T$ at each temperature $T$ was typically chosen such that $\Delta T\,\sim\,0.01\,T$. The finite oscillation amplitude in the ac-modulation technique typically causes that the specific heat feature of sharp first-order transitions, as is the case for the MI transition in $\kappa$-Mn, are slightly broadened.

\begin{figure}[t]
\includegraphics[width=\linewidth]{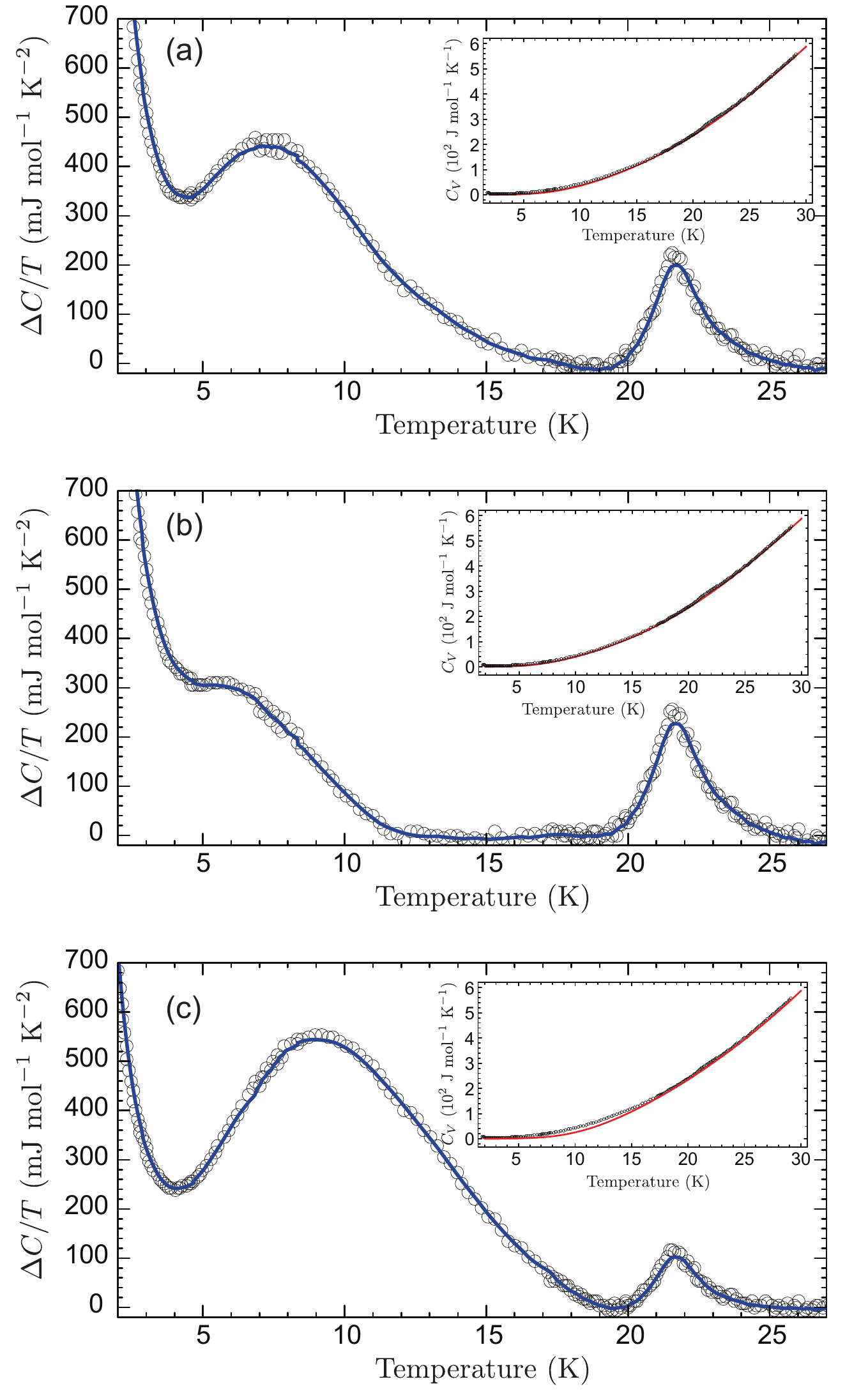}
\caption{Comparison of different background modelling for the specific heat data. In each panel (a)-(c), we show $\Delta C/T$ curves which were obtained after subtraction of different phononic background model curves. The insets in each panel show the measured raw data (open blue symbols) and the background fit (red line) that was used to obtain $\Delta C/T$ in the respective main panel. The fitting range and the fitting parameters are discussed in the text.}
\label{fig:DebyeEinstein}
\end{figure}

\textit{Background subtraction - } The specific heat of $\kappa$-Mn is dominated by phononic contributions, as is evident from the measured data shown in the insets in Fig.\,~\ref{fig:DebyeEinstein}. Unfortunately, a non-magnetic reference material is not available for an independent determination of the phononic background. This renders a precise determination of associated entropies difficult. Nevertheless, our data can be used to obtain estimates of the entropies associated with the Mn and BETS ordering. 

To this end, we obtained the anomalous contribution to the specific heat by modelling the background with a phononic contribution, $C_{v,\text{ph}}$. Below the MIT at $T_{\rm N}$, there is no charge contribution to the specific heat in $\kappa$-Mn ($\gamma\,=\,0$). Above $T_{\rm N}$, the charge contribution is finite, but likely very small compared to the phononic contribution for high $T$. Based on typical $\gamma$ values of organic charge-transfer salts \cite{Nakazawa96}, we can estimate $C_{v,\text{el}}\,=\,\gamma\,T \sim 600\,$mJ$\cdot$mol$^{-1}\cdot$K$^{-1}\,\ll\,C_{v,\text{ph}}$. As a result, we neglect electronic contributions in our background modelling.
For organic charge-transfer salts it has often been reported that the low-temperature specific heat is dominated by Debye and Einstein contributions, resulting from acoustical and low-lying optical phonons \cite{Swietlik97}. In order to keep the number of fitting parameters small, we considered only one Einstein and one Debye temperature.  Thus, we employed following Einstein-Debye form to model $C_{v,\text{ph}}$:

\begin{align}
C_{v,\text{ph}} = & \ 9k_B n_D \left(\frac{T}{\Theta_D}\right)^3 \int_0^\frac{\Theta_D}{T}\ dx \ \frac{x^4 e^x}{(e^x-1)^2}\nonumber \\
& \ +3k_B n_E \left(\frac{\Theta_E}{T}\right)^2 \frac{e^{\Theta_E/T}}{(e^{\Theta_E/T}-1)^2}
\end{align}
where $T$ is the temperature, $k_B$ is the Boltzmann constant, $\Theta_D$ is the Debye temperature, and $\Theta_E$ is the Einstein temperature. The number of phonon modes of Debye and Einstein type are given by $3n_D$ and $3n_E$, respectively, where we constrain $(n_D+n_E)/N_A = 68$, the total number of atoms per formula unit. We note that we included a scaling factor in our model in order to account for errors in the determination of the very small mass of the crystal as well as for potential errors resulting from the subtraction of addenda contributions. 
Given that there are huge magnetic contributions to the specific heat at low temperatures due to the Mn spins, we had to exclude the data at very low temperatures from the fit. Instead, we performed fits across different windows at intermediate temperatures below, but close to $T_{\rm N}$. In addition, we included the specific heat data for $T\,\geq\,24\,$K (i.e., $T\,>\,T_{\rm N}$) in each fit, since we expect this data to be largely dominated by phononic contributions.

In Fig.\,\ref{fig:DebyeEinstein} we compare the results of different background fits. In (a) we show the fit that was used in the main text. This fit (see red line in the inset) was obtained by simultaneously fitting the experimental specific heat (see open symbols in the inset) in the ranges 16\,K$\,\leq\,T\,\leq\,$19\,K and $T\,\geq\,24\,$K. The fit parameters were  $n_E = 1.8\times 10^{24}$\,mol$^{-1}$, $\Theta_E = (55\,\pm\,4)$ K, and $\Theta_D = (219\,\pm\,6)$ K. The values of $\Theta_D$ and $\Theta_E$ are well consistent with values typically observed in organic charge-transfer salts. $\Theta_D$ values typically range from 180\,K to 220\,K \cite{Wosnitza94, Katsumoto88, Andraka89, Andraka89, Steward86}. $\Theta_E$ values have been reported to be $\sim$\,28\,K for $\kappa$-(ET)$_2$I$_3$ \cite{Wosnitza94} or $\sim$\,46\,K for $\kappa$-(ET)$_2$Hg(SCN)$_2$Cl \cite{Gati18}.
Thus, we infer that this modeling of the phononic background contribution for $\kappa$-Mn is reasonable.

For comparison, we show in (b) and (c) other fits, where the lower-$T$ fitting range was varied to a larger window [11\,K$\,\leq\,T\,\leq\,$19\,K (b)] and a smaller window [19\,K$\,\leq\,T\,\leq\,$20\,K (c)]. As we will discuss now, the choice of fitting window primarily affects the amount of entropy assigned to the Mn features below $T_{\rm N}$.

The fit in (b) yielded $\Theta_D = (209\,\pm\,2)$\,K, $\Theta_E = (47\,\pm\,3$)\,K and $n_E = 1.3\times 10^{24}$ mol$^{-1}$, which are similar to the values of the fit in (a). The entropies that can be inferred from this fit amount to 
$\Delta S_{\rm Mn}$\,=\,$\int_{2}^{19} \Delta C/T \ dT \approx 2.7$ J mol$^{-1}$ K$^{-1}$ for the Mn correlations and $\Delta S_{\rm MIT}$\,=\,$\int_{2}^{19} \Delta C/T \ dT\approx 0.5$\,J mol$^{-1}$ K$^{-1}$ for the BETS ordering. Thus, shifting the fitting window does not result in a significantly different estimate of $\Delta S_{\rm MIT}$, but reduces the estimate of $\Delta S_{\rm Mn}$. The latter is not surprising, since the extended fitting window cuts off large amounts of the low $T$ entropy. Nevertheless, the low-$T$ entropy is so large that it is only reasonable to associate it with the Mn spins. 

Finally, the fit in (c) resulted in $\Theta_D = (293\,\pm\, 47)$\,K, $\Theta_E\,=\,(74\,\pm\,3)\,$K, which are clearly out of the range of reported values for organic charge-transfer salts, and $n_E = 2.7\times 10^{24}$ mol$^{-1}$. Given the small range of fitting, this phononic model can be expected to be the least accurate. Nonetheless, similar entropy estimates of $\Delta S_{\rm Mn}\,=\,\int_{2}^{19} \Delta C/T \ dT\approx 5.511$ J mol$^{-1}$K$^{-1}$ and $\Delta S_{\rm MIT}\,=\,\int_{19}^{25} \Delta C/T \ dT \approx 0.202$ J mol$^{-1}$ K$^{-1}$ were obtained. 
Thus, the conclusion that the entropy change at $T_{\rm N}$ is far too small to indicate significant coupling between the Mn and BETS spins remains robust against the broad details of the phononic background model.


\textit{Independent crosscheck of $\Delta S_{\rm MIT}$ -} In order to confirm our entropy estimate $\Delta S_{\rm MIT}$ independently, we also calculated the entropy from the Clausius-Clapeyron equation using data of the thermal expansion \cite{Agarmani21} and the published pressure dependence of $T_{\rm N}$ \cite{PhysRevB.82.155123}. This analysis yielded $\Delta S_{\rm MIT}\,\sim\,0.4\,$J$\cdot$mol$^{-1}\cdot$K$^{-1}$ which is consistent with the value inferred from specific heat in the main text. This not only confirms that our background determination of the specific heat is solid, but also that the entropy change across the MIT transition is too small for Mn atoms to be significantly involved.

\clearpage

\end{document}